\documentclass[%
 reprint,
superscriptaddress,
 amsmath,amssymb,
 aps,twocolumn
]{revtex4}

\DeclareMathOperator{\sign}{sgn}

\usepackage{graphicx}
\usepackage{dcolumn}
\usepackage{bm}
\usepackage{xcolor}
\usepackage{enumitem}

\begin{document}

\preprint{APS/123-QED}

\title{
Symmetries in 3D photoelectron momentum spectroscopy as precursory methods for dichroic and enantiosensitive measurements}

\author{Michael Davino}
\affiliation{%
 Department of Physics, University of Connecticut, Storrs, CT 06268, USA
}
\author{Edward McManus}
\affiliation{%
 Department of Physics, University of Connecticut, Storrs, CT 06268, USA
}
\author{Tobias Saule}
\affiliation{%
 Department of Physics, University of Connecticut, Storrs, CT 06268, USA
}
\author{Phi-Hung Tran}
\affiliation{%
 Department of Physics, University of Connecticut, Storrs, CT 06268, USA
}
\author{Andr\'es F. Ord\'o\~{n}ez}
\affiliation{%
 Department of Physics, Imperial College London,
SW7 2AZ London, UK
}
\affiliation{%
 Department of Chemistry, Queen Mary University of London,
E1 4NS London, UK
}
\author{George Gibson}
\affiliation{%
 Department of Physics, University of Connecticut, Storrs, CT 06268, USA
}
\author{Anh-Thu Le}
\affiliation{%
 Department of Physics, University of Connecticut, Storrs, CT 06268, USA
}
\author{Carlos A. Trallero-Herrero}
\affiliation{%
 Department of Physics, University of Connecticut, Storrs, CT 06268, USA
}
\email{carlos.trallero@uconn.edu}

\begin{abstract}
3D photoelectron angular distributions (PADs) are measured from an atomic target ionized by ultrafast, elliptical fields of opposite handedness. Comparing these PADs to one another and to numeric simulations, a difficult to avoid systematic error in their orientation is identified and subsequently corrected by imposing the dichroic symmetry by which they are necessarily related. We show that this correction can be directly applied to molecular targets in the same fields. This paves the way for measurement of enantiosensitive information which has yet to be accessed experimentally.
\end{abstract}

                              
\maketitle


Chirality has been, and continues to be, of significant interest to both fundamental science and technological applications. While biological homochirality is among the most profound unsolved scientific questions \cite{meierhenrich2013amino,sallembien2022possible}, chiral molecules are already fundamental for drug production \cite{lin2011chiral} and serve as disease biomarkers for cancer, Alzheimer's, diabetes and more \cite{liu2023detection}. Despite the need for reliable methods, chiral recognition remains challenging \cite{brandt2017added,liu2023detection}. Photoelectron momentum spectrosopy techniques present a promising alternative at both the fundamental level \cite{wanie2024capturing} and the applied level \cite{comby2023fast,comby2018real}. Specifically, photoelectron angular distributions (PADs) obtained from ionization of chiral molecules by circular or elliptically polarized strong fields not only contain information pertaining to ionization dynamics \cite{eckle2008attosecond,trabert2021nonadiabatic,hockett2015maximum}, but also for the identification of chiral enantiomers \cite{beaulieu2016universality,comby2018real}.


The interplay between field handedness and molecular chirality in ionization manifests as asymmetries in the measured PAD. Such measurements are highly relevant as they can be both dichroic (opposite for opposite field handedness) and enantiosensitive (opposite for opposite enantiomers); a particularly powerful example of this is photoelectron circular dichroism (PECD) \cite{ritchie1976theory,beaulieu2016universality}. PECD manifests as a forward-backward asymmetry in the PAD along the axis of laser propagation for randomly oriented chiral molecules when ionized by circular or elliptical fields and can be tracked on ultrafast time scales \cite{beaulieu2017attosecond,ayuso2022ultrafast,beaulieu2016probing,comby2016relaxation}. The forward-backward asymmetry of PECD does not require measurement of the full 3D momentum distribution to resolve; the 2D projection as measured by conventional velocity map imaging (VMI) techniques is generally sufficient. However, information regarding molecular chirality is not exclusively contained within this asymmetry. Theoretical work has shown that the azimuthal (in the plane transverse to field propagation) dependence of the PAD, inaccessible to 2D VMI, contains information about molecular chirality \cite{ordonez2022disentangling,demekhin2019photoelectron} which has not been accessed experimentally. Further, although standard reconstruction methods have typically allowed for 3D information to be recovered from 2D measurements in cylindrical symmetric systems, recent work suggests that symmetry breaking is expected to occur in any case where linear and circular fields are used together \cite{sparling2023importance}. Such cases are particularly relevant for studying chiral molecules.

In this letter we provide analytic methods foundational for identifying dichroism and enantiosensitivity in experimentally obtained 3D PADs. Using a novel 3D VMI setup \cite{davino2023plano} we measure strong field ionization of atomic xenon (Xe) by elliptically polarized light of different handedness. The measured 3D PADs are dichroic and non cylindrically symmetric. They are found to obey the expected dichroic symmetry rules, but only after correction for systematic errors in the azimuthal orientation. Further, we show that this correction may be directly applied to a molecular species (oxygen, O$_{2}$) ionized by the same fields. Thus our methodology may be broadly applied and is particularly relevant for molecular targets where the numeric simulations required to calibrate the correction may be infeasible. We argue this is a necessary precursor to studies involving chiral targets where azimuthal orientation is critical to the extraction of maximum enantiosensitive information.


The natural representation of 3D PADs is in the spherical harmonics,
\begin{equation}
S(k,\theta,\varphi)=\sum_{L,M}\beta_{L,M}(k)Y_{L,M}(\theta,\varphi),
    \label{eq:signal}
\end{equation}
where $k \propto |\vec{p}|$, $\theta$ is the angle from the optical propagation axis, $\hat{z}$, and the azimuthal angle, $\varphi$, is in the plane of polarization. The coefficients, $\beta_{L,M}(k)$, are the anisotropy parameters defined in 3D. The angle $\varphi$ is measured relative to the field polarization, namely, the ellipse major axis which is aligned to $\hat{x}$. Common VMI methods utilize reconstruction techniques which require a cylindrically symmetric system and therefore only $M=0$ coefficients are accessible. By performing direct 3D measurements, no assumption of the PAD symmetry is required, and therefore complementary enansiosensitive information will be contained within the $M\neq0$ terms \cite{ordonez2022disentangling}. The information lost to conventional VMI measurements becomes increasingly relevant in the case of elliptically polarized light and in the multi-photon regime where many $M$ terms are allowed.

\begin{figure*}
\centering
\includegraphics[width=0.8\linewidth]{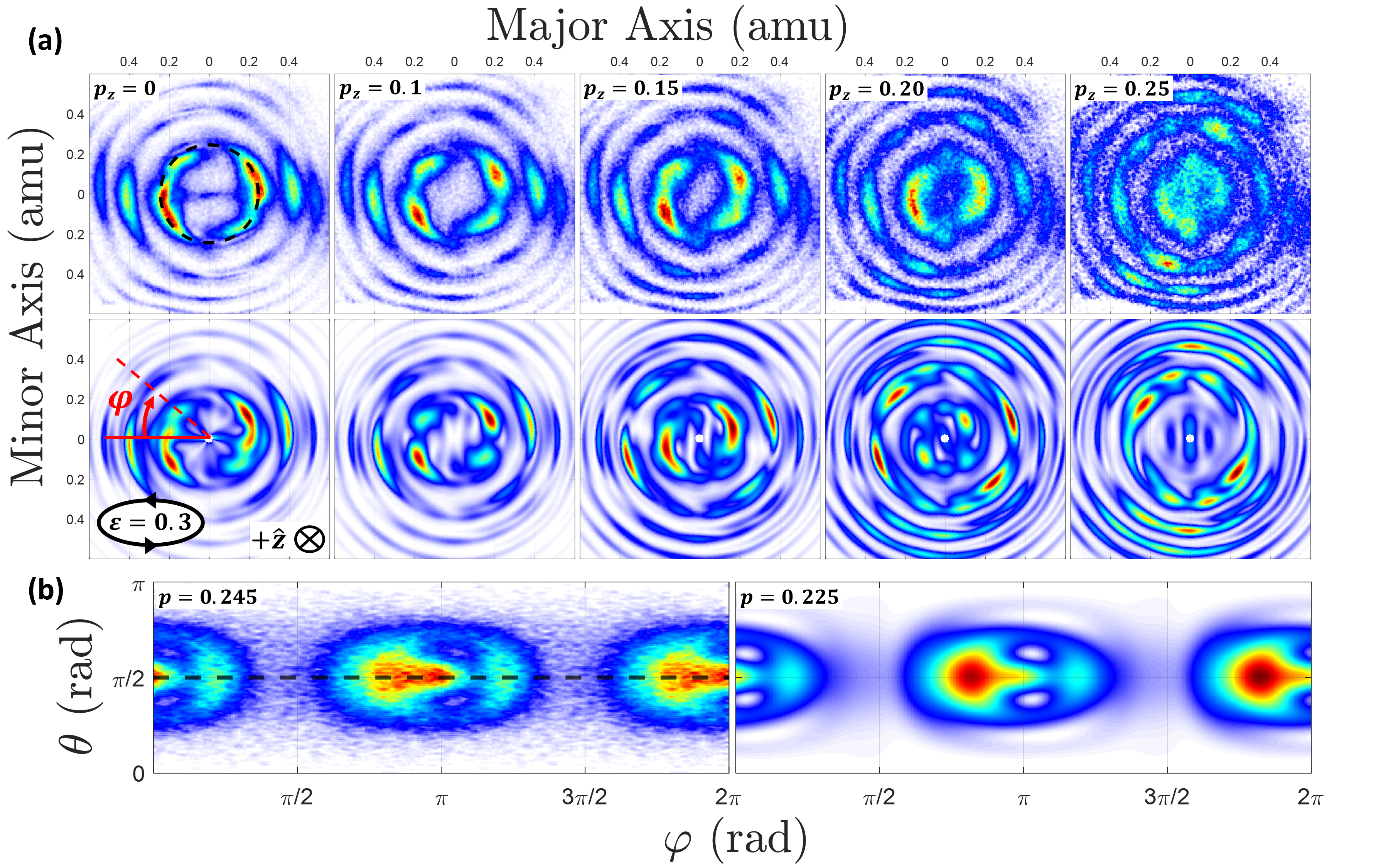}
\caption{Experimental and simulated results for Xe$^{+}$. (a) Slices of the 3D photoelectron momentum distribution in the plane of polarization for various $p_{z}$; experiment and simulation are represented in the top and bottom rows respectively. Experimental distributions are integrated over a range $\pm0.05$ amu from the quoted $p_{z}$. (b) Spherical PADs of the first ATI ring for experiment (left) and simulation (right). Each is generated from the maximum of their first ATI peak, the experimental peak at $p=0.245$ amu (integrated over $\pm0.007$ amu) and simulation at $p=0.225$. Dashed black lines in (a) and (b) represent same line, placed at the peak of the first ATI shell for $\theta=\pi/2$. All distributions are separately normalized for easy visualization of structure.}
\label{fig:exp_sim_compare}
\end{figure*}

The information encoded in PADs is neatly encapsulated in $\beta_{L,M}$ which are subject to selection rules dictated by the target symmetry \cite{ordonez2022disentangling,demekhin2019photoelectron}. Several important instances are highlighted here for the case of elliptically polarized light with ellipticity $\varepsilon=|E_{min}|/|E_{max}|$.  The field handedness will be denoted as right ($r$) or left ($\ell$) handed, and molecular handedness will be denoted as positive ($+$) or negative ($-$). Also, because any measured distribution is necessarily real, it is advantageous to use the real spherical harmonics, thus the anisotropy parameters are also real.
\begin{enumerate}[label=\roman*.$\>$]
    \item $\beta_{L,M}=0$ for odd $M$
\end{enumerate}
    This is a direct consequence of the two-fold symmetry of the system in the plane of polarization.
\begin{enumerate}[resume,label=\roman*.$\>$]
    \item Enansiosensitive coefficients have odd $L$ 
\end{enumerate}
    This condition, familiar from the forward-backward asymmetry in PECD, stems from considering a mirror reflection in the polarization plane. Such reflection exchanges enantiomers without modifying the rotation direction of the field, which leads to the relation
    \begin{equation}
        S^{+}(k,\theta,\varphi)=S^{-}(k,\pi-\theta,\varphi),
    \label{eq:enantiosensitive_relation}
    \end{equation}
    which holds for both field handedness.
\begin{enumerate}[resume,label=\roman*.$\>$]
    \item Dichroic coefficients have either odd $L$ and $M\geq0$ \underline{or} even $L$ and $M<0$
\end{enumerate}
    This condition stems from considering a 180 deg rotation of the system about the $x-$axis. Such reflection reverses the rotation direction of the field without modifying the enantiomers, which leads to the relation
    \begin{equation}
        S^{(r)}(k,\theta,\varphi)=S^{(\ell)}(k,\pi-\theta,-\varphi),
    \label{eq:chiral_dichroic_relation}
    \end{equation}
    which holds for both enantiomers. For an achiral sample, mirror symmetry about the polarization plane modifies Eq. \ref{eq:chiral_dichroic_relation} to be
    \begin{equation}
        S^{(r)}(k,\theta,\varphi)=S^{(\ell)}(k,\theta,-\varphi).
    \label{eq:achiral_dichroic_relation}
    \end{equation}
    This is the relevant dichroic relation for the work presented in this letter and has the direct consequence that $\beta_{L,M}=0$ for odd $L$.
    
Finally, for dichroic PADs satisfying either Eq. \ref{eq:chiral_dichroic_relation} or \ref{eq:achiral_dichroic_relation}, the $\varphi$-inversion manifests in the anisotropy parameters as
\begin{equation}
    \beta_{L,M}^{(r)}=
    \begin{cases}
        \beta_{L,M}^{(\ell)}, & M\geq0 \\
        -\beta_{L,M}^{(\ell)}, & M<0.
    \end{cases}
\label{eq:beta_sym}
\end{equation}

\begin{figure}
\centering
\includegraphics[width=0.45\textwidth]{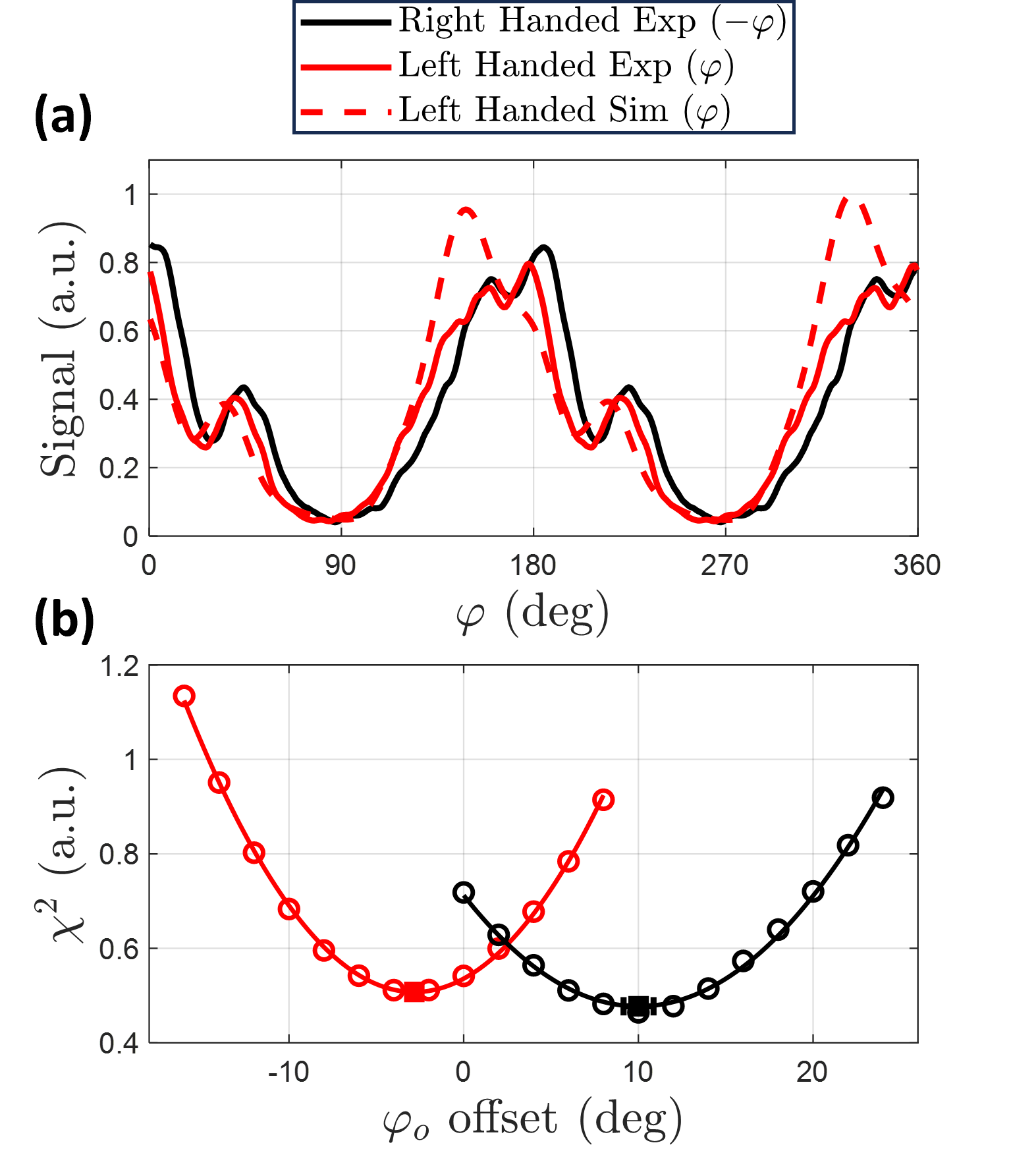}
\caption{(a) Photoelectron yield as a function of $\varphi$ for the peak of the first ATI ring at $p_{z}=0$, or equivalently, $\theta=\pi$ (see dashed black lines in Fig. \ref{fig:exp_sim_compare}). This is shown for left (experiment and simulation) and right (experimental only) handed polarization with $\varepsilon=0.3$. The right handed results are $\varphi$-inverted for direct comparison. (b) $\chi^{2}$ as a function of an introduced variable angular offset, $\varphi_{o}$. This is shown for the $S^{(\ell)}$ minimized to the simulated PAD (black), and the $S^{(r)}$ minimized to the corrected $S^{(\ell)}$ (red). The curves indicate fits to a quadratic function, and the minima are shown with errorbars.}
\label{fig:chi}
\end{figure}
    
The rules and relations presented in points i-iii and Eqs. \ref{eq:enantiosensitive_relation}-\ref{eq:beta_sym} represent strict conditions for the anisotropy parameters under various configurations of the field and target handedness. From an experimental perspective, a failure to observe these relations when they are relevant is most easily attributed to imperfect field characterization with respect to the laboratory frame. This is to say that the handedness, ellipticity, and ellipse orientation of the field must be well known in the interaction region to transform the PAD into the interaction frame which is defined by the ellipse axes. It is in this interaction frame that, for achiral targets,  Eqs. \ref{eq:achiral_dichroic_relation} and \ref{eq:beta_sym} must hold. We find that despite careful characterization of the fields outside the interaction region, comparison of the measured PADs to Eq. \ref{eq:achiral_dichroic_relation} show clear deviations from the strict selection rules. This error can attributed to birefringence of optical windows and is difficult to avoid. We show that with the aid of high-quality simulations, an angular offset in $\varphi$ may be introduced to properly align the experimental ellipse axes. The resulting PADs exhibit the expected dichroic relations Eqs. \ref{eq:achiral_dichroic_relation} and \ref{eq:beta_sym} to high precision.



The VMI apparatus used here is a double-sided electron VMI and ion time-of-flight spectrometer capable of measuring 3D PADs in coincidence with ions \cite{davino2023plano}. A Titanium:Sapphire laser system was used to generate elliptically polarized laser pulses ($80$ fs, $47$ \textmu J). These pulses were focused into the VMI chamber containing the target gas (Xe or O$_{2}$), reaching an ionizing intensity of $\sim2\times10^{13}$ W/cm$^{2}$. The ellipticity of the field and orientation of the ellipse axes were determined just prior to the VMI entrance window by the relative throughput power of the pulse following a wire grid polarizer for various angles of the polarizer. Generally, this measurement has a precision of $\pm 0.01$ and $\pm 0.5$ deg for the ellipticity and axes orientation respectively. The results shown in this letter are for $\varepsilon=0.3$, but all findings were similarly shown for $\varepsilon=0.6$. For futher details on the experimental setup, ellipticity characterization, and results for $\varepsilon=0.6$ please see Supplemental Materials.

Coincidence analysis is performed to isolate photoelectrons from Xe$^{+}$ or O$_{2}^{+}$, and from these electron events 3D PADs are generated. $\beta_{L,M}$ are extracted individually for specific $k$ and thus, from the full 3D PAD, angular histograms $h_{ij}=h(\theta_{i},\varphi_{j})$ are generated. These histograms are effectively 3D PADs of spherical 'shells' for a given radius $k$, and they will be addressed as such. The histograms are normalized such that $\sum_{L,M}|\beta_{L,M}|^{2}$ approaches $1$ with the number of terms summed. In all experiments we compute terms up to and including $L=10$ yielding sums $>0.990$. The uncertainty of each anisotropy parameter ($\sigma_{L,M}$) was determined by a data partitioning scheme of the gated electron events, and is taken to be the standard deviation of the set of $\beta_{L,M}$ as attained from the (10) different partitioned subsets. Here we focus on spherical shells with $k$ corresponding to the first above threshold ionization (ATI) peak as they contain the most structure and largest number of electron events.

To simulate the experimental results, we solved the time-dependent Schr\"odinger equation (TDSE) numerically for the target atom in an intense laser pulse. Our method has been described previously (see, e.g., \cite{hoang2017retrieval,chen2009quantitative}). Briefly, the target atom is treated within the single-active-electron (SAE) approximation with a model potential $V(r)$ of five independent parameters. More details can be found in Supplementary Materials. From the theory point of view, one of the most important findings is that measured 3D PADs provide very strict constraints on the model parameters.

\begin{figure*}[t!]
\centering
\includegraphics[width=\linewidth]{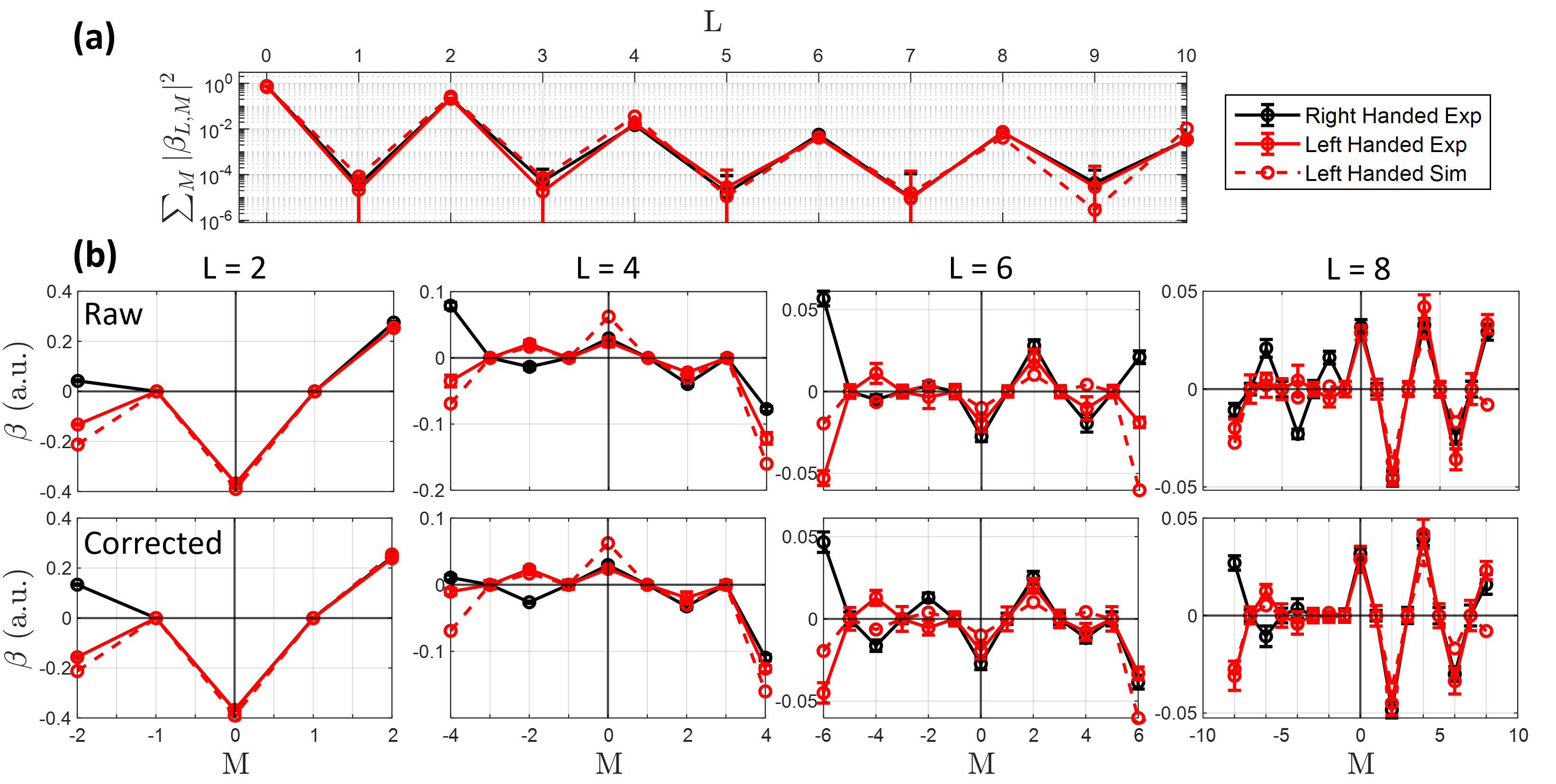}
\caption{$\beta_{L,M}$ decomposition up to $L=10$ from the first ATI shell of Xe$^{+}$ for ionization by elliptically polarized fields ($\varepsilon=0.3$) of both right and left handedness, as well as simulation results for the left handed case. (a)  Log plot of the overall contribution of each $L$; this is shown for the corrected ($\chi^{2}$ minimized) experimental PADs, although they are effectively unchanged upon correction. Error bars extending beyond the limit of the $y$-axis indicate they reach into negative values. (b) Plots of individual $\beta_{L,M}$ for $L=2,4,6,8$. Experimental results are shown before (top row) and after (bottom row) angular correction. $\beta_{L,M}$ are discrete parameters, connecting line segments are merely a visual tool.}
\label{fig:beta}
\end{figure*}


Fig. \ref{fig:exp_sim_compare} sample PADs for both experiment and simulation corresponding to ionization by left-handed fields. Shown in (a) are slices of the 3D distribution in the plane of polarization for different momenta along the laser propagation axis, $p_{z}$, and (b) presents sample spherical PADs corresponding to the first ATI peak. Qualitatively, both types of PAD exhibit exceptionally good agreement between experiment and simulation. Even the slice distributions at $p_{z}=0.25$ amu, where the ATI structure is complicated and experimental statistics are worse, there remain striking similarities. Also note that differences in experiment versus simulation may be partially attributed to the finite integration of the 3D PAD in the experimental case.

Fig. \ref{fig:chi} (a) presents lineouts of the photoelectron yield as a function of $\varphi$ for the first ATI ring at $p_{z}=0$ (equivalently, the first ATI shell at $\theta=\pi$, see dashed black lines in Fig. \ref{fig:exp_sim_compare}). The right handed results have been $\varphi$-inverted so that, by Eq. \ref{eq:achiral_dichroic_relation}, it is expected the three lineouts are identical. While the general structure of the angular photoelectron yield is very comparable between each dataset, there is a clear offset in $\varphi$ between each of the three. In particular, the two experimental lineouts are offset in $\varphi$ by $>5$ deg. This is despite measurement of the ellipse axes (and thus the PAD axes) to an estimated accuracy of $\pm 0.5$ deg. However, we reiterate that this measurement is performed prior to the VMI entrance window and we atribute this difference to birefringence in the window under vacuum stress. Regardless of the reason for the error in characterization, a poorly calibrated ellipticity at the interaction region translates as a poor transformation from the lab frame to the interaction frame.

To properly transform each experimental PAD ($S^{(\ell)}$ and $S^{(r)}$) into the interaction frame, a two step procedure is introduced. First, an angular offset  $\varphi\rightarrow\varphi+\varphi_{o}$ is introduced to $S^{(\ell)}$, to optimize the agreement between the measurement and simulation. In doing so $S^{(\ell)}$ is aligned to the field ellipse axes. Then, $S^{(r)}$ can be aligned in one of two equivalent ways. One approach is to simply repeat the process of comparing to simulation. Alternatively, and the method used here, Eq. \ref{eq:achiral_dichroic_relation} may be employed as an optimization condition where the newly corrected $S^{(\ell)}$ serves as a reference to which $S^{(r)}$ is aligned by its own angular offset. Note that in this approach, if the experimental PAD used as reference is not itself properly aligned, Eq. \ref{eq:achiral_dichroic_relation} will still hold, but neither of the two PADs will be in the interaction frame nor will their field axes coincide. Done properly, the net result is two experimentally obtained PADs, from opposite field handedness, in the interaction frame.


Optimization of the agreement between two PADs, whether it be experiment to simulation or experiment to experiment, may be performed by conventional $\chi^{2}$ minimization with respect to $\varphi_{o}$. The details can be found in Supplemental Materials. Fig. \ref{fig:chi} (b) presents $\chi^{2}$ minimization for the first ATI shell utilizing left handed simulation as a reference to orient the left (red) and right (black) handed experimental results. Each $\chi^{2}$ is minimized within $\pm1$ deg; this represents a significant improvement in ellipse alignment considering the right handed results require a correction of $\sim10$ deg.

The consequences of this correction procedure on the anisotropy parameters in Fig. \ref{fig:beta} (a) shows the relative contribution of each $L$ to the full PAD normalized as stated above. Fig. \ref{fig:beta} (b) presents both the raw and corrected $\beta_{L,M}$ for the first four even $L\neq0$.

\begin{figure}[t!]
\centering
\includegraphics[width=\linewidth]{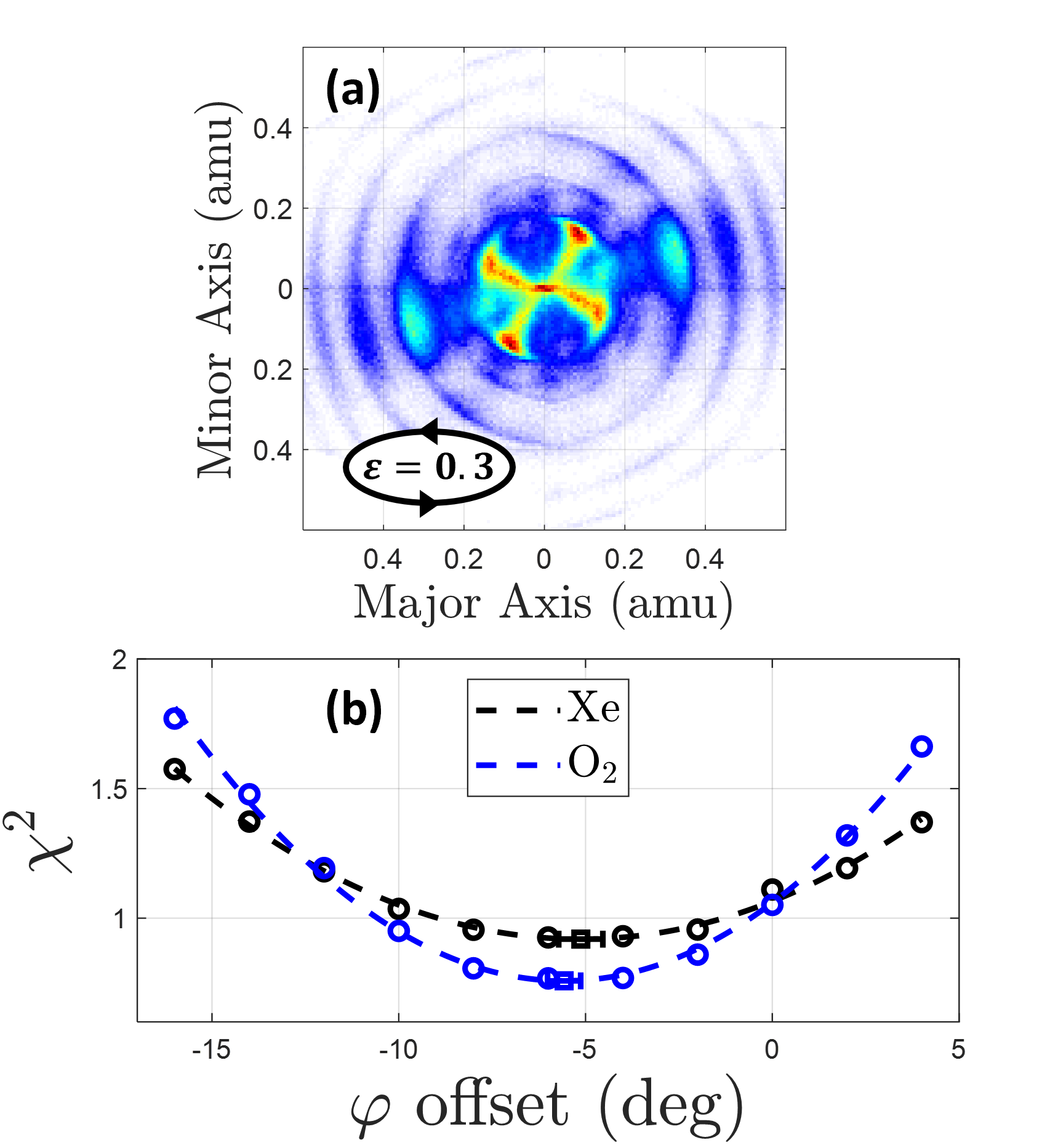}
\caption{(a) Slice of the 3D PAD for O$_{2}$ ionized by field ellipticity $\varepsilon=0.3$. The slice is centered at $p_{z}=0$ and integrated over a width of $\pm0.05$ amu. (b) $\chi^{2}$ minimization for Xe and O$_{2}$ ionized by the same fields. Dashed lines indicate fits to a quadratic function, and the offset angles for which they are minimized fall within the uncertainty of one another.}
\label{fig:species}
\end{figure}

Inspection of Fig. \ref{fig:beta} reveals that, only after $\chi^{2}$ correction, each of the expected selection rules and relations for an achiral target are well represented. Most obvious is that $\beta_{L,M}=0$ lies within the error of each parameter with odd $M$ and odd $L$. Further, it can be seen qualitatively that the relation between coefficients from opposite handedness (Eq. \ref{eq:beta_sym}) holds very well up to $L=10$. To assess the agreement quantitatively we introduce a weighted mean square deviation from Eq. \ref{eq:beta_sym},
\begin{equation}
    \eta^{2}=\frac{1}{N}\sum_{L,M}\frac{|\beta^{(\ell)}_{L,M}-\sign \left[ M \right]\beta^{(r)}_{L,M}|^{2}}{|\sigma^{(\ell)}_{L,M}|^{2}+|\sigma^{(r)}_{L,M}|^{2}},
\end{equation}
where $N$ is the number of summation terms, i.e. the number of $L,M$ combinations included. The factor of $\sign \left[ M \right]$ is necessary to account for Eq. \ref{eq:beta_sym}, that is, the right handed coefficients are expected to be equal but opposite their left handed counterparts for $M<0$. Thus the numerator is minimized for every $L$ and $M$ combination when Eq. \ref{eq:beta_sym} is satisfied. $\eta^{2}<1$ indicates that, on average, the anisotropy parameters satisfy Eq. \ref{eq:beta_sym} to within their uncertainties. The anisotropy parameters shown in Fig. \ref{fig:beta} result in $\eta^{2}=4.09$ for the uncorrected PADs, and $\eta^{2}=0.31$ for the corrected PADs. These values represent a significant shift in the interpretation of results upon angular correction: the original PADs would be said to not exhibit a dichroic relation whereas the corrected PADs do. Furthermore, we need to emphasize that for more complex PADs the weight of different $L$ components can not be determined \textit{a priori} and thus a cutoff in the expansion is very target dependent.

Next, we extend the same analysis to molecular targets to demonstrate the generality of the azimuthal calibration. 3D PADs were measured for both Xe and O$_{2}$ for both field handedness. A sample of the O$_{2}$ PAD at $p_{z}=0$ is presented in Fig. \ref{fig:species} (a). For both species, an angular offset was introduced to $S^{(r)}$ and $\chi^{2}$ minimized to optimize Eq. \ref{eq:achiral_dichroic_relation}; the results are presented in Fig. \ref{fig:species} (b).  Note that the same field parameters ($80$ fs, $47$ \textmu J, $\varepsilon=0.3$) are used here as have been used throughout this letter, but $\chi^{2}$ minimization was performed without first orienting to simulation; thus the discrepancy in the Xe angular offset between Figs. \ref{fig:chi} and \ref{fig:species}. The critical result is that both Xe and O$_{2}$ have a minimum $\chi^2$ for the same offset angle. In other words, the correction angle maximizing the dichroic relation (Eq. \ref{eq:beta_sym}) for Xe, also does so for O$_{2}$. This means that when simulations are utilized to find offset angles which properly orient Xe in the interaction frame, the same offset angles may be applied directly to other molecules, including chiral molecules, for which such simulations cannot be done.

To conclude, we have shown the experimental capability to measure 3D PADs that exhibit dichroic behavior. Since the field can not be characterized at the interaction region, we find systematic errors that affect the lab-to-interaction frame transformation. This transform is critical in extracting dichroic and enantiosensitive information contained within the anisotropy parameters that represent the PADs. As a correction, numeric simulations are utilized in conjunction with the symmetry relations between PADs. In particular, the dichroic relation is shown to be well satisfied with the correction up to the $L=10$ term in the expansion. We extend the characterization to O$_{2}$ and demonstrate it shares the same azimuthal correction angle as Xe, implying that the interaction frame transformation may be calibrated in an atomic species and then applied to a molecular species where numeric simulations are infeasible. This work paves the way for high-resolution measurement of the enantiosensitive information contained within the anisotropy parameters by providing a contrast of several orders of magnitude between dichroic only (even L) and enantiosensitive (even and odd L) values in $\beta_{L,M}$. Further, 3D PADs present theory with highly sensitive experimental information for model validation without the assumption of any symmetry.

Experimental work was done under Air  Force  Office  of  Scientific  Research grant FA9550-21-1-0387. Theoretical simulations were done by PT and A-TL under DOE BES, Chemical Sciences, Geosciences, \& Biosciences Division grant DE-SC0023192. Data analysis carried out by EM supported under DOE BES, Chemical Sciences, Geosciences, \& Biosciences Division grant DE-SC0024508.

\bibliography{references}

\end{document}


\preprint{APS/123-QED}

\title{Supplementary Materials}

\author{Michael Davino}
\affiliation{%
 Department of Physics, University of Connecticut, Storrs, CT 06268, USA
}
\author{Tobias Saule}
\affiliation{%
 Department of Physics, University of Connecticut, Storrs, CT 06268, USA
}
\author{Edward McManus}
\affiliation{%
 Department of Physics, University of Connecticut, Storrs, CT 06268, USA
}
\author{Phi-Hung Tran}
\affiliation{%
 Department of Physics, University of Connecticut, Storrs, CT 06268, USA
}
\author{Andr\'es F. Ord\'o\~{n}ez}
\affiliation{%
 Department of Physics, Imperial College London,
SW7 2AZ London, UK
}
\affiliation{%
 Department of Chemistry, Queen Mary University of London,
E1 4NS London, UK
}
\author{George Gibson}
\affiliation{%
 Department of Physics, University of Connecticut, Storrs, CT 06268, USA
}
\author{Anh-Thu Le}
\affiliation{%
 Department of Physics, University of Connecticut, Storrs, CT 06268, USA
}
\author{Carlos A. Trallero-Herrero}
\affiliation{%
 Department of Physics, University of Connecticut, Storrs, CT 06268, USA
}
\email{carlos.trallero@uconn.edu}

\maketitle


\section{Experimental Setup}

\begin{figure}[h]
\centering
\includegraphics[width=0.4\textwidth]{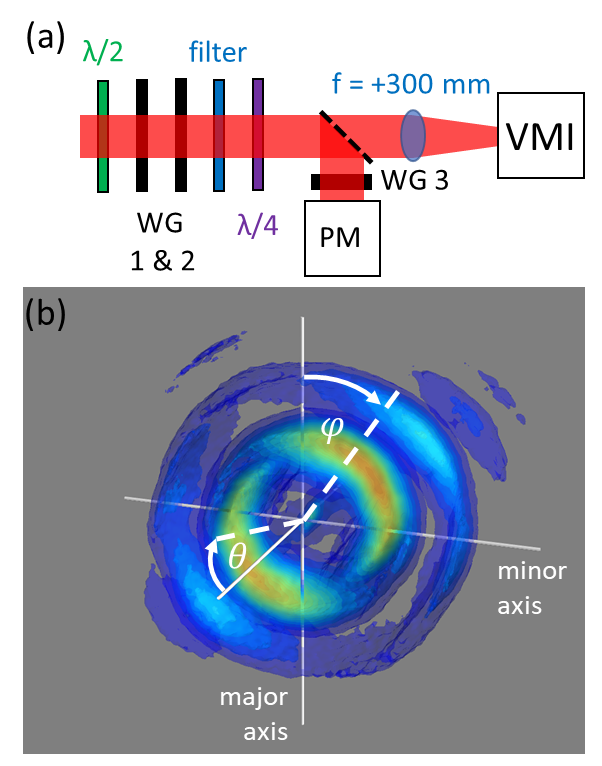}
\caption{(a) Depiction of the optical setup incident on the VMI. Elements labelled WG denote wire grid polarizers, the dashed line indicates a removable mirror, and PM being the power meter. The filter is a band-pass filter of width 10 nm centered at 800 nm. (b) A sample 3D PAD for $\varepsilon=0.3$; note that the unlabeled axis corresponds to the axis of laser propagation.}
\label{fig:setup}
\end{figure}

A depiction of the experimental setup is shown in Fig. \ref{fig:setup} (a). The output of a Titanium:Sapphire laser system is incident on a $\lambda/$2 achromatic wave plate and two wire grid polarizers which serve to orient and clean the pulse polarization. The laser pulse is subsequently incident on a 10 nm spectral filter centered at 800 nm and followed by a $\lambda/$4 achromatic waveplate which controls the ellipticity of the pulse polarization. The ellipticity ($\varepsilon=|E_{min}|^{2}/|E_{max}|^{2}$) and orientation of the ellipse axes are determined by the relative power throughput of the pulse following a wire grid polarizer for various angles of the polarizer. Generally, this measurement has a precision of $\pm 0.01$ and $\pm 0.5$ deg for the ellipticity and axes orientation respectively; see Section \ref{ellipticity_char}. The final pulse was measured to have an intensity autocorrelation profile with a full-width at half-maximum (FWHM) of $\sim$80 fs, pulse energy of 47 $\mu$J, and is focused into the VMI to a peak focal intensity of $\sim$2 x 10$^{13}$ W/cm$^{2}$. The 3D VMI apparatus used here is a double-sided electron VMI and coincident ion TOF spectrometer; the VMI side is a plano-convex thick-lens design.

\section{Theoretical calculation details}
\label{sect:theory}
As mentioned, the target atom is treated within the single-active-electron (SAE) approximation. The model potential $V(r)$, as a function of five independent parameters, is constructed with the self-interaction free density functional theory. The potential parameters are optimized by fitting the calculated binding energies to the experimental binding energies of the ground state and the first few excited states of the target atom. The initial wave function is propagated in time using the split-operator method to obtain the final wave function at the end of the laser pulse. The photoelectron momentum distribution is then calculated by projecting the final wave function onto the eigenstates of the continuum electron. For Xe atom, all contributions from $5p$ electrons with $m=0$ and $\pm 1$ were included. To compare with experimental measurements, we also carried out the interaction volume averaging, assuming that the laser pulse is a Gaussian beam. The laser parameters were chosen to closely match the experimental values. To save computer time, we have limited ourselves to the pulse duration (the FWHM) of about $10$ fs. To avoid artificial reflections due to a finite box size, we chose a relatively large box with a typical $r_{max}=1000$ a.u. All the parameters were carefully tested to ensure the numerical results converged.

\section{Ellipticity Characterization}
\label{ellipticity_char}

\begin{figure*}[htb!]
\centering
\includegraphics[width=0.8\linewidth]{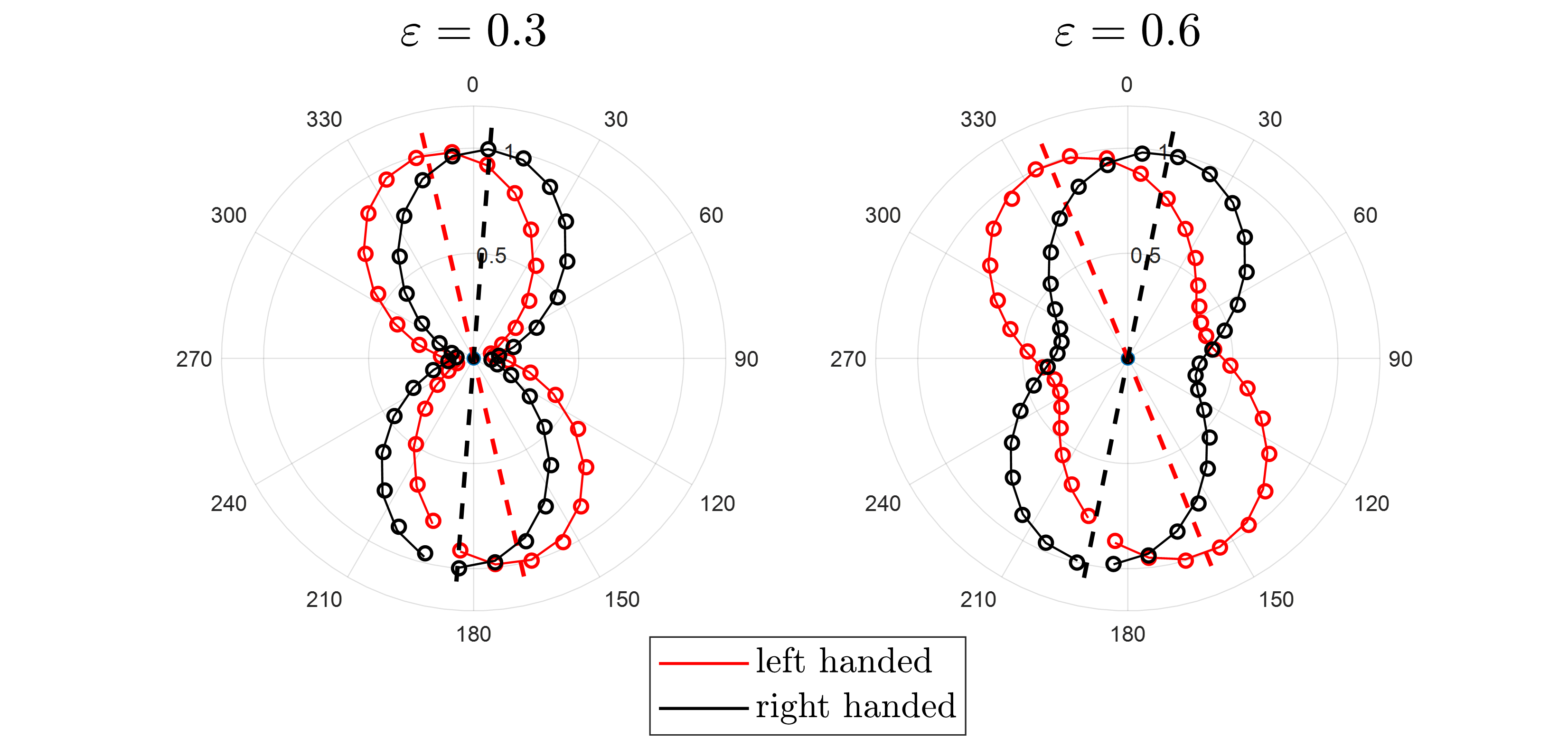}
\caption{Polar plots depicting the throughput power dependence on the angle of the fast axis of a wire grid polarizer (WG3 in the main text) for $\varepsilon=0.3$ and $0.6$; the zero angle corresponds to the VMI axis. Markers represent the measured throughput power, solid lines represent cosine-squared fits, and dashed lines represent the ellipse major axis extracted from the each fit.}
\label{fig:ellipticity}
\end{figure*}

The ellipticity and orientation of the fields are determined by the relative throughput power of the laser pulse following a wire grid polarizer for various angles of the polarizer; polar plots of this dependence are shown in Fig. \ref{fig:ellipticity}. Fitting throughput power as a function of polarizer angle to cosine-squared yields the orientation and ellipticities quoted in Tab. \ref{tab:ellipticity}. The key point here is that according to these measurements of the field, the orientation of the ellipse axes should be known to $< 0.35$ deg. This is shown to not be the case in the main text.

\begin{table}[ht]
\centering
\begin{tabular}{|l|l|l|}
\hline
Handedness & Ellipticity ($\varepsilon$) & Major Axis (deg) \\
\hline
\hline
Left & 0.293$\pm$0.009 & -13.01$\pm$0.27 \\
\hline
Right & 0.298$\pm$0.007 & 4.44$\pm$0.22 \\
\hline
\hline
Left & 0.604$\pm$0.004 & -21.95$\pm$0.31 \\
\hline
Right & 0.577$\pm$0.003 & 11.31$\pm$0.24 \\
\hline
\end{tabular}
\caption{\label{tab:ellipticity} All values of ellipticity and ellipse major axis orientation (angle with respect to VMI axis) as attained from cosine-squared fits show in Fig. \ref{fig:ellipticity}.}
\end{table}

\section{Experimental Data Analysis}

For a given experimental dataset, electron events occurring in a laser shot with coincidence measurement of a particular ion (Xe$^{+}$ or O$_{2}^{+}$) are compiled by ($x,y,t$) and converted into spherical momentum representation ($k,\theta,\varphi$). As we must extract $\beta$ for particular $k$ individually, electron events are gated on a narrow range about some $k$ of interest, we will be concerned with $k$ corresponding to the first above-threshold ionization (ATI) peak as it contains the most structure and largest number of electron events.

Photoelectron angular distributions (PADs) are constructed as 2D histograms, $h_{ij}=h(\theta_{i},\varphi_{j})$, from the gated electron events. In doing so we impose the condition that $h_{ij}$ be two-fold rotationally symmetric in $\varphi$ to reduce experimental irregularities. Note that doing so forces the odd $M$ anisotropy parameters closer to zero. That said, even without imposing this symmetry, $\beta_{L,M}=0$ is within uncertainty for all odd $M$ as is claimed in the main text. The symmetry is imposed by by having each electron event contribute to $h_{ij}$ twice, once at the measured $\varphi$, then again at $\varphi+\pi$. The uncertainty of each histogram bin is taken to be $\Delta h_{ij}=\sqrt{h_{ij}}$. The histograms are then normalized to have a unit integral, and the uncertainties are scaled accordingly. Note that simulated results are analyzed to produce PADs of the same dimension as their experimental counterparts.

For projecting the PADs onto the spherical harmonics we note that physically they represent the angular probability distributions for an ejected electron in momentum space for a specific $k$, i.e. $|\Psi_{k}(\theta,\varphi)|^{2}$. This being the case, it is more natural to project $\sqrt{h_{ij}}$ onto the spherical harmonics as to reconstruct $|\Psi_{k}|$. Doing so has the convenient consequence that $\sum_{L,M}|\beta_{L,M}|^{2}\rightarrow1$ with the number of terms summed.

Uncertainty in $\beta_{L,M}$ (denoted $\sigma_{L,M}$) was determined by randomly partitioning the collection of electron events contributing to a given PAD into 10 subsets. From each subset $h_{ij}$ was generated and $\beta_{L,M}$ extracted. $\sigma_{L,M}$ was then taken to be the standard deviation of the set of $\beta_{L,M}$ as generated from each subset.

To correct the PAD alignment, $\chi^{2}$ is minimized with respect to $\varphi_{o}$, as defined by
\begin{equation}
    \chi^{2}(\varphi_{o})=\frac{1}{N}\sum_{i,j}\frac{|h^{exp}_{ij}(\varphi_{o})-h^{sim}_{ij}|^{2}}{|\Delta h^{exp}_{ij}(\varphi_{o})|^{2}}
    \label{eq:chi}
\end{equation}
where $N$ is the number of summation terms, i.e. histogram bins. In the case of minimizing $\chi^{2}$ between two experimental histograms ($h^{(\ell)}_{ij}$ and $h^{(r)}_{ij}$) the expression becomes
\begin{equation}
    \chi^{2}(\varphi_{o})=\frac{1}{N}\sum_{i,j}\frac{|h^{(r)}_{ij}(\varphi_{o})-h^{(\ell)}_{ij}|^{2}}{|\Delta h^{(r)}_{ij}(\varphi_{o})|^{2}+|\Delta h^{(\ell)}_{ij}|^{2}}.
\end{equation}
Note that this expression is used where one of the two histograms has been $\varphi-$inverted such that $\chi^{2}$ is minimized when the dichroic relation is satisfied.

\section{Results for $\varepsilon$ = 0.6}

\begin{figure}[t!]
\centering
\includegraphics[width=0.8\linewidth]{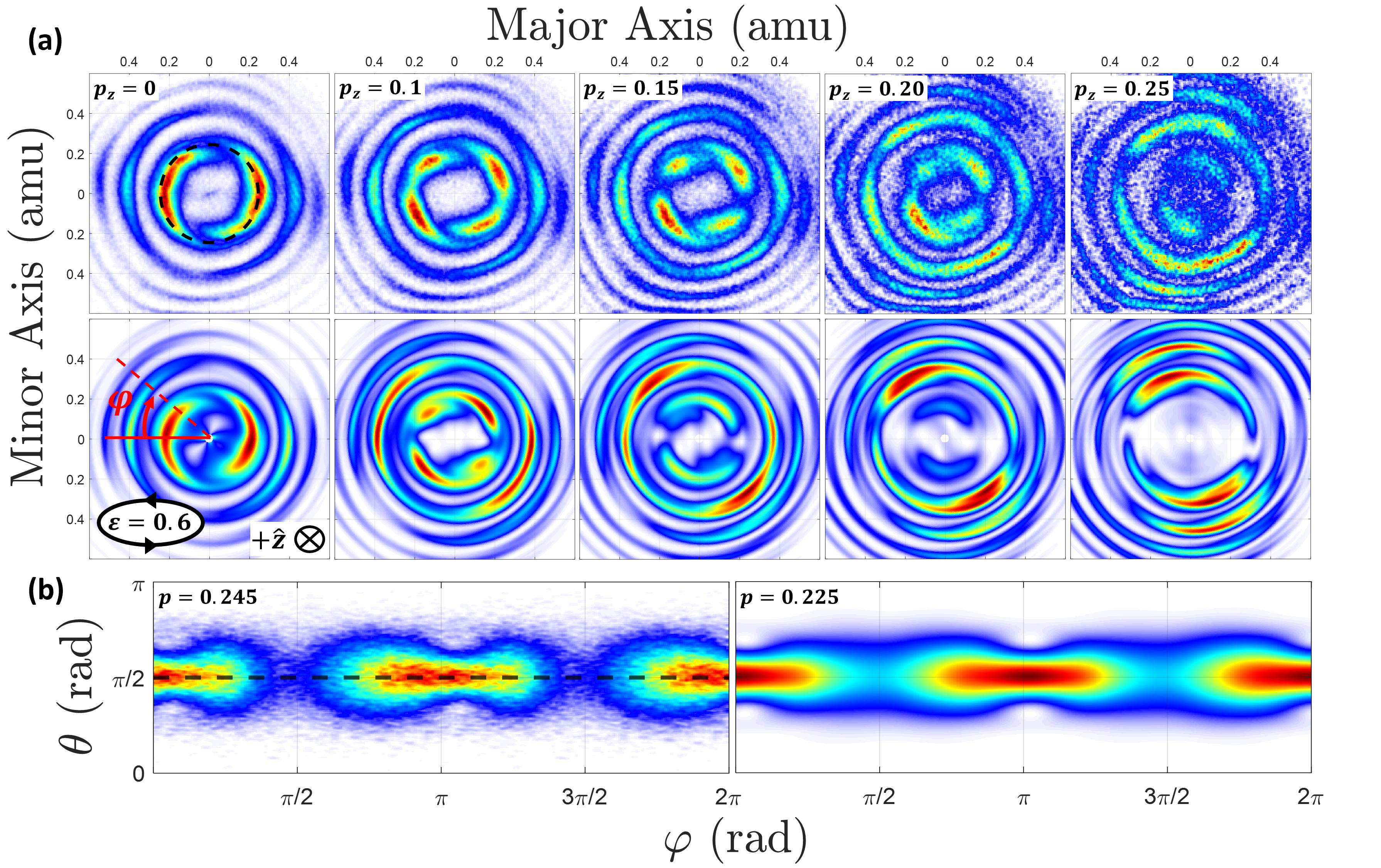}
\caption{Experimental and simulated results for Xe$^{+}$ ionized by elliptically polarized laser pulses. (a) Slices of the 3D photoelectron momentum distribution in the plane of polarization for various $p_{z}$; experiment and simulation are represented in the top and bottom rows respectively. Experimental distributions are integrated over a range $\pm0.05$ amu from the quoted $p_{z}$. (b) Spherical PADs of the first ATI ring for experiment (left) and simulation (right). Both are generated at the maximum of their first ATI peak, the experimental peak at $p=0.245$ amu (integrated over $\pm0.007$ amu) and simulation at $p=0.225$. Dashed black lines in (a) and (b) represent same line, placed at the peak of the first ATI shell for $\theta=\pi/2$. All distributions are self normalized for easy visualization of structure.}
\label{fig:exp_sim_compare2}
\end{figure}

\begin{figure}[b!]
\centering
\includegraphics[width=0.8\textwidth]{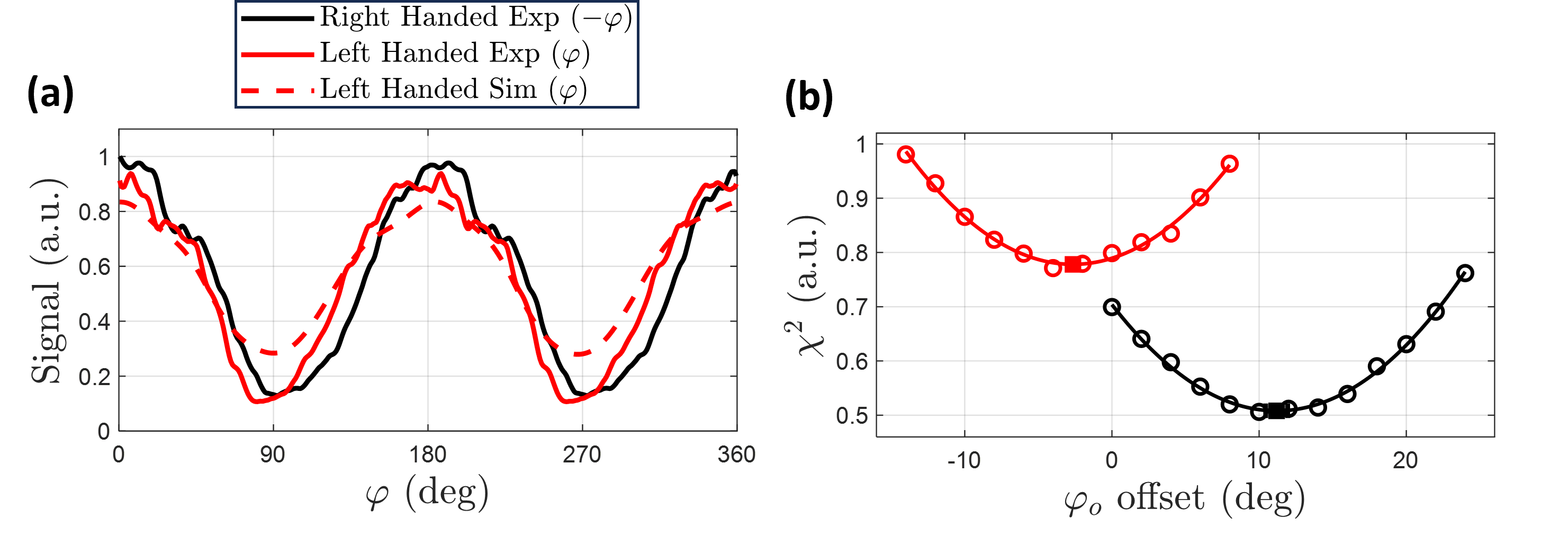}
\caption{(a) Photoelectron yield as a function of $\varphi$ for the peak of the first ATI ring at $p_{z}=0$, or equivalently, $\theta=\pi$ (see dashed black lines in Figure \ref{fig:exp_sim_compare2}). This is shown for right (experiment and simulation) and left handed (experimental only) polarization. The left handed results are plotted as $-\varphi$ for direct comparison. (b) $\chi^{2}$ as a function of an introduced variable angular offset, $\varphi_{o}$. Dashed lines indicate fits to a quadratic function, and the minima are shown with errorbars.}
\label{fig:chi2}
\end{figure}

\begin{figure}[b!]
\centering
\includegraphics[width=\linewidth]{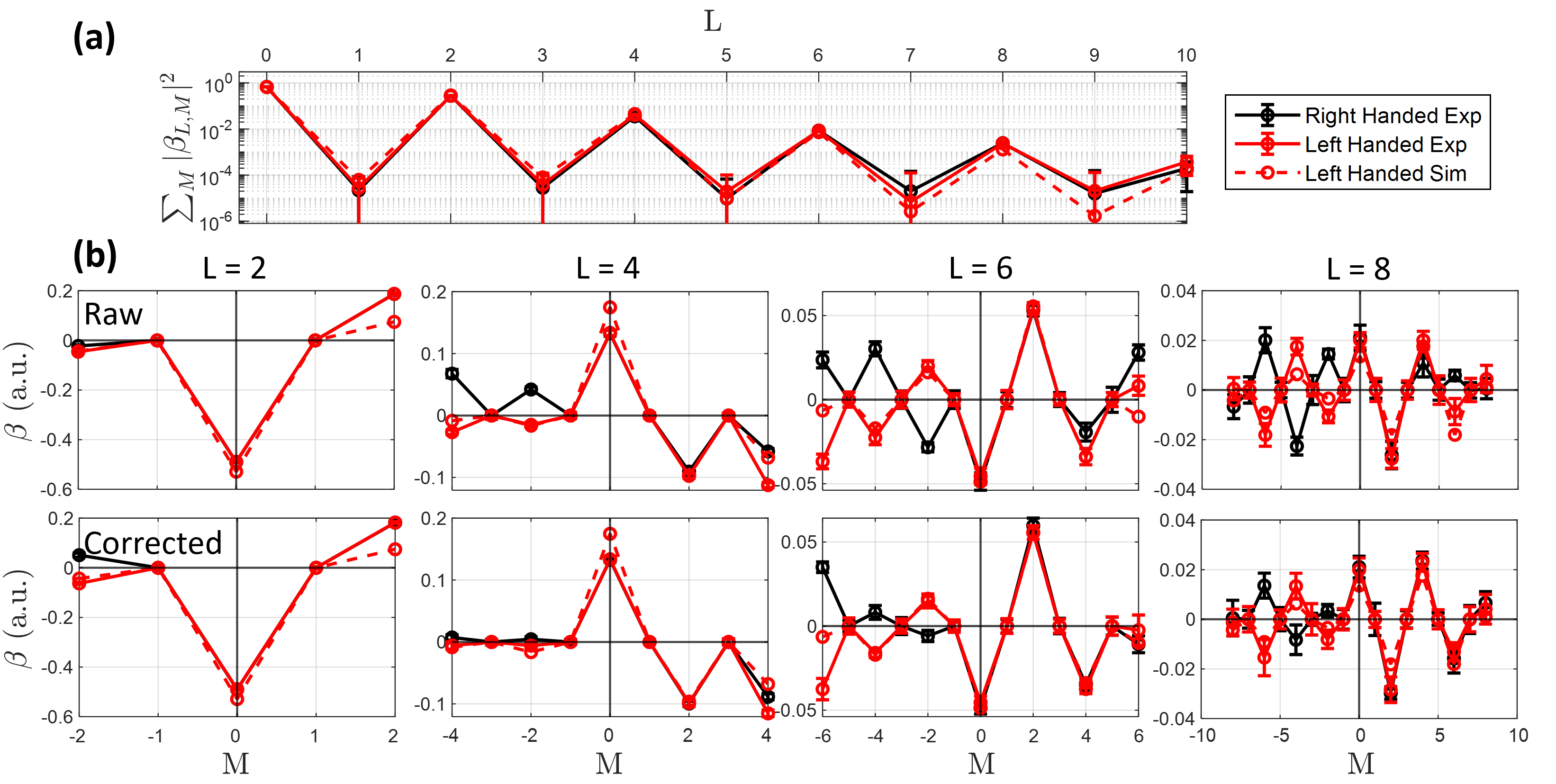}
\caption{$\beta_{L,M}$ decomposition up to $L=10$ of PAD shells corresponding to the first ATI ring from Xe$^{+}$ for ionization by elliptically polarized fields of both right and left handedness, as well as simulation results for the right handed case. Right handed experimental and simulated results have been shifted by $\varphi=8$ and $3$ deg respectively as to minimize $\chi^{2}$. Shown is the contribution of each $L$ (a) and the real and imaginary components of $\beta_{L,M}$ for all parameters $L=2,4,6,8$ (b). Be mindful that $\beta$ are discrete parameters, connecting line segments in each plot are merely a visual tool.}
\label{fig:beta2}
\end{figure}

Supplementary Figs. \ref{fig:exp_sim_compare2} - \ref{fig:beta2} are exactly comparable to Figs. 1 - 3 in the main text but with $\varepsilon=0.6$. All general findings from these figures are exactly as those described in the main text. Figure captions and discussions from the main text are repurposed here for convenience.
\\
\\

Fig. \ref{fig:exp_sim_compare2} (b) shows sample spherical PADs, corresponding to the first ATI peak, generated for both experiment and simulation of ionization by left-handed fields. Shown in (a) are slices of the 3D distribution along the axis of propagation. Qualitatively, both types of PAD exhibit good agreement between experiment and simulation. Note that differences in experiment versus simulation may be partially attributed to the finite integration of the 3D PAD in the experimental case. That said, although a relatively large integration width is used here for easy visualization of the distributions, the general structure is largely consistent with smaller widths.

The PADs from Fig. \ref{fig:exp_sim_compare2} are compared to the PAD measured under the same conditions but produced from a field of opposite (right) handedness. Fig. \ref{fig:chi2} (a) presents lineouts of the photoelectron yield as a function of $\varphi$ for the first ATI ring at $p_{z}=0$ (equivalently, the first ATI shell at $\theta=\pi$, see dashed black lines in Fig. \ref{fig:exp_sim_compare2}). The right handed results have been $\varphi$-inverted so that it is expected the three lineouts are identical. While the general structure of the angular photoelectron yield is very comparable between each dataset, there is a clear offset in $\varphi$ between each of the three.

To optimize the agreement between the various spherical PADs, $\chi^{2}$ minimization is performed with respect to $\varphi_{o}$. Fig. \ref{fig:chi2} (b) presents $\chi^{2}$ minimization for the first ATI shell utilizing left handed simulation as a reference to orient the left (red) and right (black) handed results. Each $\chi^{2}$ is minimized within $\pm1$ deg; this represents a significant improvement in ellipse alignment.

The consequences of this correction procedure on the anisotropy parameters is presented in Fig. \ref{fig:beta2}. (a) shows the relative contribution of each $L$ to the full PAD; each of the three sets has $\sum^{L=10}_{L,M}|\beta_{L,M}|^{2}>0.990$. Fig. \ref{fig:beta2} (b) presents both the raw and corrected $\beta_{L,M}$ for the first four even $L\neq0$.

Inspection of Fig. \ref{fig:beta2} finds that, after $\chi^{2}$ correction, each of the expected selection rules and relations for an achiral target are well represented. Most obvious is that $\beta_{L,M}=0$ lies within the error of each parameter with odd $M$ and odd $L$. The anisotropy parameters shown in Fig. \ref{fig:beta2} result in $\eta^{2}=2.05$ for the uncorrected PADs, and $\eta^{2}=0.25$ for the corrected PADs. These values represent a significant shift in the interpretation of results upon angular correction: the original PADs would be said to not exhibit a dichroic relation whereas the corrected PADs do.